\documentclass[aip,reprint]{revtex4-2}

\usepackage{graphicx}
\usepackage{booktabs}
\usepackage{amsmath}
\usepackage{amssymb}
\usepackage[normalem]{ulem}
\usepackage[T1]{fontenc}
\usepackage{hyperref}

\usepackage{xcolor}
\newcommand{\LS}[1]{\textcolor{green}{#1}}
\newcommand{\LSA}[1]{\textcolor{green}{#1}}
\newcommand{\CC}[1]{\textcolor{red}{#1}}
\newcommand{\FF}[1]{\textcolor{blue}{#1}}
\newcommand{\PG}[1]{\textcolor{cyan}{#1}}
\renewcommand{\LS}[1]{#1}
\renewcommand{\LSA}[1]{#1}
\renewcommand{\CC}[1]{#1}
\renewcommand{\FF}[1]{#1}
\renewcommand{\PG}[1]{#1}

\begin{document}

\title{End-to-End Analysis of Charge Stability Diagrams with Transformers}

\author{Rahul Marchand}
\thanks{Shared first authorship}
\affiliation{Department of Engineering Science, University of Oxford, Oxford OX1 3PJ, United Kingdom}

\author{Lucas Schorling}
\thanks{Shared first authorship}
\affiliation{Department of Engineering Science, University of Oxford, Oxford OX1 3PJ, United Kingdom}

\author{Cornelius Carlsson}
\affiliation{Department of Engineering Science, University of Oxford, Oxford OX1 3PJ, United Kingdom}

\author{Jonas Schuff}
\affiliation{Department of Materials, University of Oxford, Oxford OX1 3PH, United Kingdom}
\affiliation{Intel Corporation, Technology Research Group, Hillsboro, OR 97124, USA}

\author{Barnaby van Straaten}
\affiliation{QuTech and Kavli Institute of Nanoscience, Delft University of Technology, P.O. Box 5046, Delft, 2600 GA, Delft, The Netherlands}

\author{Taylor L. Patti}
\affiliation{NVIDIA Corporation, 2788 San Tomas Expressway, Santa Clara, 95051, CA, USA}

\author{Federico Fedele}
\affiliation{Department of Engineering Science, University of Oxford, Oxford OX1 3PJ, United Kingdom}

\author{Joshua Ziegler}
\affiliation{Intel Corporation, Technology Research Group, Hillsboro, OR 97124, USA}

\author{\CC{Parth Girdhar}}
\affiliation{Department of Engineering Science, University of Oxford, Oxford OX1 3PJ, United Kingdom}

\author{Pranav Vaidhyanathan}
\thanks{Corresponding author: pranav.vaidhyanathan@stcatz.ox.ac.uk}
\affiliation{Department of Engineering Science, University of Oxford, Oxford OX1 3PJ, United Kingdom}

\author{Natalia Ares}
\thanks{Corresponding author: natalia.ares@eng.ox.ac.uk}
\affiliation{Department of Engineering Science, University of Oxford, Oxford OX1 3PJ, United Kingdom}

\date{\today}

\begin{abstract}
Transformer models and end-to-end learning frameworks are rapidly revolutionizing the field of artificial intelligence. In this work, we apply \LS{object detection transformers} to analyze charge stability diagrams in semiconductor quantum dot arrays, a key task for achieving scalability with spin-based quantum computing. 
\LS{Specifically, our model} identifies triple points and their connectivity, which is crucial for virtual gate calibration, charge state initialization, drift correction, and pulse sequencing.
We show that it surpasses Convolutional Neural Networks in performance on three different spin qubit architectures, all without the need for retraining.
In contrast to existing approaches, our method significantly reduces complexity and runtime, while enhancing generalizability.
The results highlight the potential of \LS{transformer-based end-to-end learning frameworks} as a foundation for a scalable, device- and architecture-agnostic tool for \LS{control and tuning of quantum dot devices.}

\end{abstract}

\pacs{}%

\maketitle %

\section{Introduction}
    
\FF{\LS{Transformers and other attention-based models} underpin the success of recent progress in artificial intelligence~\cite{attentionisallyouneed, transformersmetareinforcementlearners, lin2021surveytransformers, visiontransformer}. 
Initially developed for \LS{natural language processing, transformers advance previous sequential-modeling approaches such as recurrent neural networks (RNN)}, and exhibit  exceptional generalization capabilities along with high versatility \LS{~\cite{zhou2024transformersachievelengthgeneralization,Goodfellow,kamath2022, whittle2025distribution}}. 
This enabled them to also outperform convolutional (CNN) \LS{on image-related tasks \cite{visiontransformer}} and to be rapidly adopted across a range of areas, including genomics \LS{\cite{roy2024alphafold3}}, robotics~\LS{\cite{sanghai2024advancestransformersroboticapplications}}, and more recently, quantum technologies~\cite{vaidhyanathan2024quantumfeedbackcontroltransformer}.}

In this work, we adapt a transformer-based model to analyze charge stability diagrams (CSD) of semiconductor quantum devices. Efficient feature extraction from CSDs is crucial for tuning and operating quantum dot arrays for quantum computing applications.
Within quantum technologies, spin qubits based on gate-defined semiconductor quantum dots represent one of the leading platforms for realizing high-fidelity~\cite{gateabovecodethreshold,gatesabovethreshold2}, hot \LS{($\gtrsim$ 1 K)}~\cite{hotqubits,hotqubits2}, and scalable quantum processors~\cite{neyens2024probing, zwerver2022qubits, burkard2023semiconductor}. However, as the number of devices scales~\cite{fedele2021, philips2022}, the large number of parameters that must be finely optimized for each individual qubit presents a major bottleneck to scalability. In this context, automated control strategies, particularly those based on machine learning, are emerging as essential tools to enable reliable and efficient device operations~\cite{Ares2021, nvidiaAIforquantumreview}.

\FF{Previous work relied on diverse machine learning techniques for automating tasks such as double quantum dot tuning~\cite{Severin_2024, bridingthegap, vanstraaten2022rfbasedtuningalgorithmquantum, RLefficientmeasurement}, multi-parameter crosstalk compensation~\cite{hickie2023}, spin-readout optimization~\cite{SchuffPSB2023}, \LS{meta-learning qubit characteristics \cite{ourmetalearning}}, and ultimately, the fully autonomous tuning of a spin qubit~\cite{schuff2024fullyautonomoustuningspin, carlsson2025automatedallrftuningspin}. }
\LS{However, previous approaches have typically relied on a variety of methods, including pre- and post-processing procedures, template matching, Hough transforms, (convolutional) neural network image or pixel classifiers, and combinations thereof~\cite{mavis, unetpixelclassifier, Mills_2019, coulombdiamondestimation, tuningchargestatesold, raybasedstateidentification, hader2024automated, wang2024efficient}.
These solutions can be time-consuming and often result in pipelines with limited generalizability.}

\begin{figure*}[ht]
    \centering
    \includegraphics[width=1.0\linewidth,]{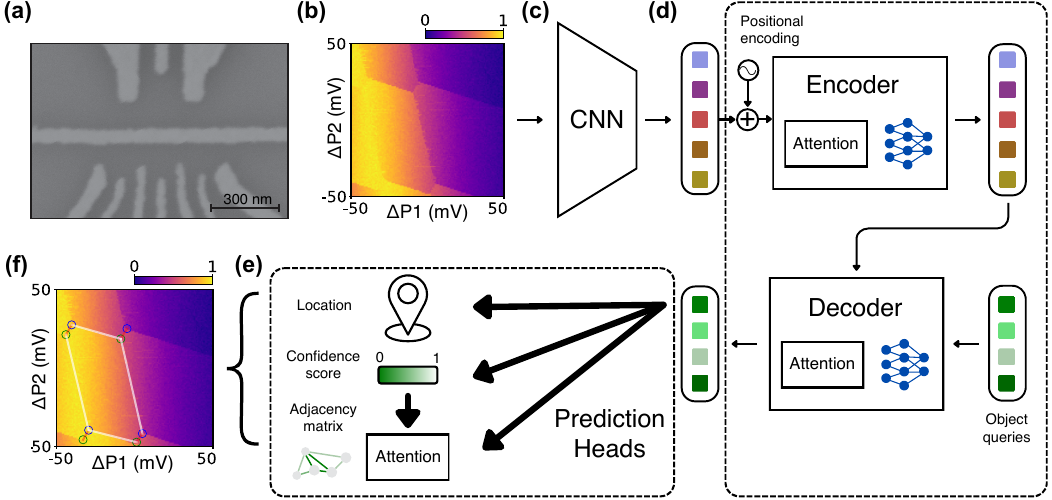}
    \caption{\textbf{Overview of TRACS' architecture}.
    \textbf{(a)} \FF{Micrograph of a depletion-mode Ge/SiGe device similar to the one used in this work (Device B). \textbf{(b)} Example of a (simulated) charge stability diagram that can be fed as an input to TRACS. \textbf{(c)} The first stage of TRACS is a convolutional neural network.} After a linear mapping, the resulting pixel feature vectors serve as tokens. Tokens are enriched with positional encoding and fed into the encoder of a transformer \textbf{(d)}. Its output and a set of learnable object queries form the input to a transformer decoder. Both encoder and decoder use the attention mechanism and multilayer perceptrons as fundamental building blocks. \textbf{(e)} Prediction heads ``interpret'' the decoder's outputs and return confidence scores on whether an instance is a triple point pair, their location, and their connectivity. \textbf{(f)} The original charge stability diagram used as an input [from (b)]  is returned, overlayed with the final transformer output.}
    \label{fig:model}
\end{figure*}

\LSA{To address these challenges, we introduce TRACS (Transformers for Analyzing Charge Stability diagrams), a transformer-based model designed to process CSDs and extract their full graph structure. TRACS identifies the pattern of lines indicating charge transitions, and the points where these lines intersect (called triple points), enabling virtual gate calibration, charge state initialization, drift correction, and pulse sequencing. Furthermore, TRACS offers fast inference and robust generalization across diverse experimental conditions and device architectures, also due to its end-to-end learning paradigm \cite{wilder2020endendlearningoptimization,glasmachers2017limitsendtoendlearning}. In end-to-end learning approaches, a single trainable model maps inputs to outputs without intermediate stages.}

\FF{We evaluate TRACS' performance on triple point and charge transition detection, using experimental data acquired from three different quantum devices: an accumulation-mode Si/SiGe device \cite{george202412}, a depletion-mode Ge/SiGe heterostructure \cite{saez2024microwave}, and an accumulation-mode Ge/SiGe heterostructure \cite{john2024two, stehouwer2025exploitingepitaxialstrainedgermanium}.}
\FF{Notably, TRACS exhibits \LS{strong performance and outperforms} a popular CNN object detection architecture \cite{yolo}, achieving \CC{a median error in placement of triple point coordinates within 3\% of the voltage scan range} while remaining agnostic to device materials and gate architectures.} Our work thus opens the door to more robust, general, and scalable strategies for autonomous quantum dot tuning, characterization, and control.

\section{Methods}
\label{sec:methods}
\LS{Our algorithm, TRACS, builds on the powerful object detection transformer (DETR)\cite{DETR}, and processes the input CSD through a convolutional neural network, followed by a transformer, and finally a set of prediction heads. While CNNs excel at local feature extraction, demonstrate robustness to pixel noise, and exhibit translational invariance, they are limited in their ability to assimilate information across those local features.
Transformers \cite{attentionisallyouneed}, on the other hand, are powerful sequence-to-sequence models that can reason across all elements in a sequence, or all local features, via the attention mechanism. The attention mechanism models pairwise relations between all elements (or so-called tokens) in a dynamic manner via learned parameters. The final prediction heads map the outputs of the transformer to the final outputs.
These are the locations
of the triple points, their confidence scores, and an adjacency matrix for their connectivity (i.e., tracing of
the charge transitions). The novel prediction head for connectivity is based on the attention mechanism\cite{attentionisallyouneed}.}

\FF{The pipeline of TRACS is schematized in Fig. }\ref{fig:model}.
The model inputs are charge sensed charge stability diagrams acquired by measuring the current or rf voltage output of a proximal charge sensor, while sweeping two gate voltages of a quantum dot device.~\cite{vigneau2023probing} %
The resulting image, with a single scalar value per pixel (channel dimension 1), is normalized to the range $[0, 1]$ and provided to TRACS as input.
This image is then passed through a convolutional neural network. The core building blocks of CNNs are learnable filters, that are convolved with the image, and pooling layers, which downsample the input while increasing channel dimension. 
\FF{A linear layer is applied to each pixel of the CNN output to generate tokens, which serve as input to the transformer encoder. These tokens are high-dimensional vector representations, \LS{ over which the attention mechanism operates to model pairwise relationships.}}
Positional encoding is added to the tokens to retain spatial information since the transformer architecture is permutation-invariant. Several attention layers between the tokens (self-attention), as well as multi-layer perceptrons, transform the input tokens of the encoder into tokens with an enriched feature representation.
\begin{figure*}[ht]
    \centering
    \includegraphics[width=\linewidth]{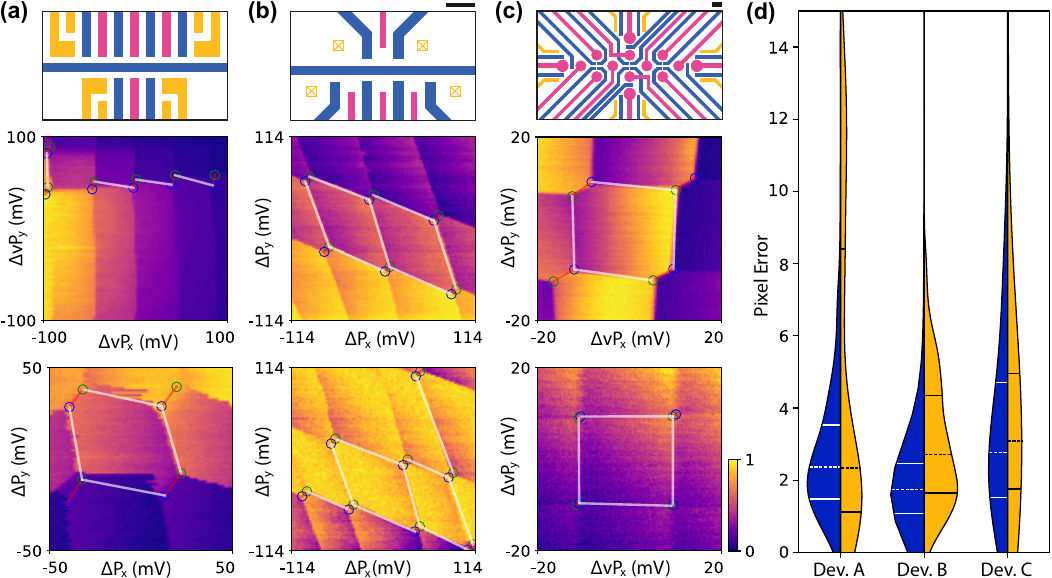}
    \caption{\textbf{Device schematics, example outputs, and triple point location performance of TRACS}. \textbf{(a)--(c)} Schematics of the devices used in this work, accompanied by two CSDs and the output of TRACS overlayed; (a) Device A: an accumulation-mode electron device on a Si/SiGe heterostructure, (b) Device B: a depletion-mode hole device on a Ge/SiGe heterostructure, (c) Device C: an accumulation-mode hole device on a Ge/SiGe heterostructure. Pink, blue and yellow layers correspond to plunger, barrier and accumulation (or ohmic) gates (or reservoirs), respectively. Additional screening layers have been omitted for clarity. The scale bar in each panel corresponds to 100\! nm, see Ref.~\cite{george202412} for details on Device A. \LS{ $\text{P}_{\text{\{x,y\}}}$ corresponds to a plunger gate and $\text{vP}_{\text{\{x,y\}}}$ corresponds to a virtualized plunger gate.} \textbf{(d)} \CC{Violin plots of the} error in triple point locations, measured as a pixel distance, of TRACS (blue) and YOLO (yellow). Dashed lines indicate median errors and solid lines separate lower and upper quartiles. 10 pixels correspond to $10/128\sim 8\%$ of the mV range for a CSD, see Appendix \ref{app:details_detr}.}
    \label{fig:results}
\end{figure*}
This encoder output is fed into the transformer decoder along with a set of learnable tokens called object queries, \LS{which can be thought of as empty slots for triple point instances.} The decoder uses a combination of cross-attention, self-attention on the object queries, and multi-layer perceptrons, to yield the final output tokens. Cross-attention contextualizes the relationship between encoder output tokens and object queries.

Each output token represents a potential triple point instance as a high-dimensional vector. Three prediction heads map the tokens to the final outputs.
The first head yields a pair of confidence scores, distinguishing a pair of triple points (both values close to 1) from no pair (both values close to 0) via a linear projection and sigmoid function. 
The second head outputs four values corresponding to the $x$ and $y$ coordinates for a pair of triple points via a linear projection and clamping.
The third head consists of two layers of attention and multi-layer perceptrons and outputs an adjacency matrix of shape $(n_{\text{obj}},n_{\text{obj}},3)$. Here, $n_{\text{obj}}$ is the number of output tokens and the three values at index $[i, j, :]$ correspond to ``probabilities''. The first ``probability'' indicates whether no line is present between token $i$ and token $j$, the second that the \LSA{upper} triple point of token $i$ is connected with token $j$, and the third that the \LSA{lower} triple point of token $i$ is connected with token $j$. Since upper triple points can only be connected to lower triple points, the matrix elements corresponding to the same connection are averaged.

TRACS is trained in a supervised manner via gradient-based optimization of the loss function
\begin{equation}
    \mathcal{L}_{\text{total}}= \mathcal{L}_{\text{BCE}}^{\text{con}}+ 
    \mathcal{L}_{\text{wing}}^{\text{loc}}+\mathcal{L}_{\text{focal}}^{\text{adj}}.
\end{equation}
The binary cross-entropy loss $\mathcal{L}_{\text{BCE}}^{\text{con}}$ penalizes wrong confidence scores. The wing loss \cite{feng2018winglossrobustfacial} $\mathcal{L}_{\text{wing}}^{\text{loc}}$ penalizes incorrect triple point locations and is more effective for computer vision localization tasks than standard Euclidean distance. The focal loss \cite{lin2018focallossdenseobject} $\mathcal{L}_{\text{focal}}^{\text{adj}}$ acts on the adjacency matrix, down-weights the contributions of well-classified examples, and takes class imbalance into account. \LS{Each of these terms is formally defined in Appendix \ref{app:details_detr}}.
 
The set of predicted triple point pairs and the ground truth set are both unordered. However, an efficient bijective matching can be computed to minimize  $\mathcal{L}_{\text{BCE}}^{\text{con}} + \mathcal{L}_{\text{wing}}^{\text{loc}}$. In practice, we normalize and weight the different terms of $\mathcal{L}_{\text{total}}$ as well as use regularization. 
See Appendix \ref{app:details_detr} for further details, and the original DETR publication \cite{DETR} for background.

The TRACS model is trained on 500,000 charge stability diagrams generated together with ground truth labels using QArray \cite{qarray, qarray_rel}. This GPU-accelerated simulator computes ground-state charge configurations of quantum dot arrays using a constant capacitance model. For closer resemblance with experimental CSDs, thermal broadening, white noise, telegraph noise and latching are included in the training data, which is supported by QArray. Simulation parameters are sampled from uniform distributions, with bounds selected to span a wide range of experimental conditions, see the \hyperref[sec:data_availability]{Data Availability}. 

Although TRACS is deliberately restricted to simulated-data training in this work, its transformer architecture supports a two-stage training paradigm \cite{devlin2019bert, hu2022lora, marouf2024mini}. In the first stage, the model undergoes pre-training on large amount of simulated data; in the second, it is ``fine-tuned'' on a smaller set of experimental data.

\section{Results}
\label{sec:results}

We test our model on CSDs acquired from three different device architectures. These are shown schematically in Fig. \ref{fig:results}(a)--(c). Device A is an accumulation-mode electron device on a Si/SiGe heterostructure, Device B is a depletion-mode hole device on a Ge/SiGe heterostructure, and Device C is an accumulation-mode hole device on a Ge/SiGe heterostructure. Refer to Appendix \ref{app:set_up} for details on the measurement set-up and to Appendix \ref{app:furtherdata} for test set sizes. For each device, two examples of CSDs are presented together with the output of TRACS. TRACS is capable of identifying triple points and their connections with high accuracy across CSDs with different noise characteristics and varying numbers of triple points. \LS{In particular, in panel (a), we showcase TRACS' capabilities on highly latched and non-centered data which, a pixel classification-based method typically fails to handle effectively. In panel (c), we highlight TRACS' capacity to operate on (over-) virtualized data, \LSA{enabling verification of the virtualization process and supporting subsequent tuning steps.}}

We benchmark TRACS against a fully CNN-based model -- the \textbf{Y}ou \textbf{O}nly \textbf{L}ook \textbf{O}nce (YOLO) architecture \cite{yolo} -- keeping the CNN from the first stage of TRACS. YOLOv1 was chosen for its efficiency, simplicity, and proven effectiveness in real-time object detection tasks. However, it does not allow for line detection, \LS{without significant modification to the original architecture.} Training and test data remain the same.

\begin{table}[ht]
    \centering
    \caption{Model evaluation and benchmark of TRACS (our model) vs. YOLOv1 (CNN-based model) on point detection (PD). \LS{Line Detection (LD) results are reported for TRACS only, since YOLO is not suited for this task.} The results include median slope error of TRACS corresponding to capacitive values for experimental data from three different device architectures as well as simulated test set.}
    \label{tab:model_performance}
    \begin{ruledtabular}
    \begin{tabular}{lcccc}
        & \textbf{Dev A} & \textbf{Dev B} & \textbf{Dev C} & \textbf{Simul} \\
        \hline
        \textbf{PD TRACS Recall}      & 0.87 & 0.84 & 0.90  & 0.98 \\
        PD YOLOv1 Recall       & 0.84 & 0.77 & 0.76  & 0.92 \\
        \textbf{PD TRACS Precision}   & 0.93 & 0.94 & 0.92  & 0.94 \\
        PD YOLOv1 Precision    & 0.86 & 0.95 & 0.85  & 0.94 \\
        \textbf{LD TRACS Recall}       & 0.95 & 0.95 & 0.91  & 0.98 \\
        LD YOLOv1 Recall       & - \footnotemark[1] & -\footnotemark[1] & -\footnotemark[1]  & - \footnotemark[1]\\
        \textbf{LD TRACS Precision}    & 0.93 & 0.85 & 0.94  & 0.94 \\
        LD YOLOv1 Precision       & -\footnotemark[1]  & - \footnotemark[1]& -  \footnotemark[1]& -\footnotemark[1] \\
        \textbf{Median slope error }         &0.031 &0.039 &0.015 &0.022\\

    \end{tabular}
    \end{ruledtabular}
    \footnotetext[1]{Original YOLOv1 architecture cannot predict line connectivity.}
\end{table}

The first four rows of Table \ref{tab:model_performance} report the recall and precision of TRACS and YOLO for triple point detection on test sets from all three devices, as well as a test set of simulated CSDs. Recall and precision formalize the notion of the ``percentage of correct predictions'' and are defined by $TP/(TP+FN)$ and $TP/(TP+FP)$. $TP$ denotes the number of true positive predictions, $FN$ the number of false negative predictions, and $FP$ the number of false positive predictions. 
For all benchmarks, we use a classification threshold of 0.5, i.e. confidence scores above this threshold are classified as triple points. Increasing the threshold improves precision, but reduces recall.
A ground truth triple point is considered correctly detected if the predicted location is within 15 image pixels of the actual position, see Appendix \ref{app:details_detr} for details. 
\LS{TRACS outperforms YOLO on the simulated test set and generalizes significantly better to \LSA{quantum device data} with an improvement of more than 5 percentage points on average for recall and precision. This corresponds to a $45 \%$ reduction in false negatives (missed triple points) and a $19 \%$ reduction in false positives.}

The distribution of error in triple point locations is shown in Fig. \ref{fig:results}(d). This is only well defined and reported for all true positive triple points, hence, the blue and yellow areas are not equal in size. For all devices, TRACS outperforms YOLO, \LS{as evidenced by its lower third-quartile bar.} 

Rows five to eight of Table \ref{tab:model_performance} report recall and precision for all connections between true positive triple points. Here, TRACS once again generalizes well to experimental CSDs with high accuracy.

From the connectivity, calculating the slope of lines between triple points becomes trivial. Slopes correspond to the strength of capacitive coupling between the gates swept in a CSD and are important quantities for their virtualization. We report a median slope error \PG{in Table 1}. 

With TRACS, point and line detection is performed in a single forward pass through the model. On an Intel core i7 CPU, we measure an average inference time of 83\! ms and expect faster inference on a GPU. This is 1-3 orders of magnitude faster than inference times reported in previous work \cite{mavis}.

\section{Conclusion \& Outlook}
\label{sec:conclusion}

In this work, we introduced TRACS for the end-to-end analysis of charge stability diagrams with transformers. \LS{TRACS automatically detects triple points in CSDs and infers their connectivity, crucial information for automated device tuning procedures, in particular for gate virtualization, navigating to charge states, and pulse sequencing. With inference in tens of milliseconds, TRACS can also be woven into closed-loop control algorithms for charge-offset drift correction.
TRACS generalizes well from simulated data to a wide range of measured CSDs, as evidenced by its strong performance across three different device architectures, including SiGe-based devices operating both in depletion and accumulation modes.}

It is also important to note that, instead of relying on specific tuning procedures, TRACS abstracts charge stability diagrams into connectivity graphs of triple points. This abstraction makes it possible to apply the same control routines across different device architectures, multi-dot arrays, and material platforms, without redesigning the underlying tuning logic.

\LS{TRACS lays the groundwork for a new generation of tuning algorithms. Next steps include the integration of more advanced experiment planners and protocols. We anticipate that our approach will become a core component of next-generation quantum dot control systems, dramatically accelerating device characterization, tuning, and ultimately the deployment of large‐scale quantum processors.}

\begin{acknowledgments}
\LS{The authors would like to thank Jaime Saez-Mollejo and Georgios Katsaros for the fabrication and measurement of Device B.} \LSA{The authors would also like to thank the authors of reference~\cite{mavis} for making their device data openly available.}
N.A. acknowledges support from the European Research Council (grant agreement 948932), the Royal Society (URF-R1-191150), Innovate UK project AutoQT (grant number 1004359) and the United States Army Research Office under Award No. W911NF-24-2-0043. Views and opinions expressed are however those of the authors only and do not necessarily reflect those of the European Union, Research Executive Agency or UK Research \& Innovation. Neither the European Union nor UK Research \& Innovation can be held responsible for them. P.V. is supported by the United States Army Research Office under Award No. W911NF-21-S-0009-2. L.S. and C.C. acknowledge support from the UKRI Doctoral Training Partnership related to EP/W524311/1 (project refs. 2886876 and 2887634). 
\end{acknowledgments}

\section*{Author Declarations}
\subsection*{Conflict of Interest}
Natalia Ares declares a competing interest as a founder of QuantrolOx, which develops machine learning-based software for quantum control. All remaining authors declare no conflicts of interest.

\subsection*{Author Contributions}
\LS{
\noindent \textbf{Rahul Marchand}: Software (lead); Data curation (lead); Formal analysis (lead); Methodology (lead).
\textbf{Lucas Schorling}: Writing – original draft (lead); Visualization (lead); Supervision (equal); Methodology (equal).
\textbf{Cornelius Carlsson}: Conceptualization (equal), Investigation (equal), Writing – review \& editing (equal); Visualization (equal); Methodology (supporting).
\textbf{Jonas Schuff}: Conceptualization (equal); Methodology (equal); Investigation (supporting).
\textbf{Barnaby van Straaten}: Investigation (equal); Software (supporting).
\textbf{Taylor L. Patti}: Methodology (supporting); Software (supporting).
\textbf{Federico Fedele}: Writing – review \& editing (equal). \textbf{Joshua Ziegler}: Investigation (equal). \textbf{Parth Girdhar}:
Supervision (supporting).

\textbf{Pranav Vaidhyanathan}: Supervision (equal): Methodology (equal);  Writing – review \& editing (equal). \textbf{Natalia Ares}: Funding acquisition (lead); Supervision (equal);  Writing – review \& editing (equal).}

\section*{Data Availability}
\label{sec:data_availability}
All test data and the software code, including files to generate the training data, will be made public upon final publication.

\appendix

\section{Details of TRACS}
\label{app:details_detr}

TRACS is trained on CSDs with a $128 \times 128$ pixel resolution. During inference, CSDs with other resolutions are linearly rescaled to match the expected dimensions. Therefore, for a specific CSD, the pixel error can be converted to millivolts (mV) by dividing it by 128 and multiplying by the CSD's voltage range in mV.

The architecture of the convolutional neural network and its parameters are similar to ResNet-18 \cite{resnet}, adapted to handle greyscale images, and work with lower image resolutions than those in the ImageNet dataset. \PG{The} first layer of filters fed to the transformer expects a channel dimension of 1, hence pooling layers as well as the final averaging and fully connected layer of ResNet-18 were removed, and strides were decreased. The modified ResNet-18 outputs $16 \times 16$ pixels with a channel width of 512. This is mapped to the transformer input of 256 tokens of dimension 128 via a convolution with kernel size 1. The transformer encoder and decoder consist of 4 layers with 8 attention heads and have a dropout of 0.1 and a feedforward dimension of 2048. 

The inputs to the connectivity prediction head are the decoder output tokens and the corresponding two confidence scores. Those confidence scores and tokens are concatenated and linearly projected to dimension 128. Those are fed into a transformer encoder architecture with 2 layers and 1 attention head.

For every CSD, we have a set of labels for pairs of triple points consisting of confidence scores $c_{i,p}\in \mathbb{R}$, locations $x_{i,p} \in \mathbb{R}^2$, where $i$ and $j$ refer to the instances of triple point pairs and $p=1,2$ refers to each point in the pair individually. Furthermore, we have three values $a_{ij,m}\in \mathbb{R}$ for the connectivity between instance $i$ and instance $j$ for $m=1,2,3$.

The TRACS predictions are given by $\hat{c}_{i,p}$, $\hat{x}_{i,p}$, and $\hat{a}_{ij,m}$. The total loss function can be written as 

\begin{align}
    \mathcal{L}_{\text{total}}&= \mathcal{L}_{\text{BCE}}^{\text{con}}+ \mathcal{L}_{\text{wing}}^{\text{loc}}+\mathcal{L}_{\text{focal}}^{\text{adj}} \notag \\
    &= \alpha_1 \sum_{i=1}^{n_q}   \sum_{p=1,2} -c_i \log{(\hat{c}_{\sigma(i)})} -\beta_1 (1-c_i) \notag \\
    &\log(1-\hat{c}_{\sigma(i)})
    \notag \\
    &+\alpha_2 \sum_{\substack{i=1\\ c_i = 1}}^{n_q}  \sum_{p=1,2} \mathcal{L}_{\text{wing}} (x_i, \hat{x}_{\sigma(i)})\notag \\
    &+ \alpha_3 \sum_{\substack{i,j=1\\ c_i = 1\\ c_j = 1}}^{n_q,n_q}  \sum_{m=1,2,3} w_m l_{\sigma(i)\sigma(j),m} (1-\hat{l}_{\sigma(i)\sigma(j),m})^{\gamma} \notag \\
    &\log(\hat{l}_{\sigma(i)\sigma(j),m}),
\end{align}
where $\alpha_1=1, \alpha_2=0.05, \alpha_3= 0.5$ weight the different loss terms, $n_q=25$ is the number of object queries, $\beta_1= 0.5$, $w_1= 0.3,w_2= w_3= 1$ are weights to counteract class imbalance, and $\gamma=2$. $\sigma(i)$ represents the permutation of labels and object tokens based on the minimal cost matching via the Hungarian algorithm. The double subscripts $c_i=1$ indicate that the sum only includes terms fulfilling this condition. $\mathcal{L}_{\text{wing}}$ has width $\omega= 0.05$ and curvature $\epsilon= 0.01$.

We use the AdamW optimizer with a learning rate of 2e-5, a learning rate scheduler, weight decay of 0.0001, batch size of 256 and train for 225 epochs.

\section{Experimental Set-up}
\label{app:set_up}

All data presented in this paper was acquired via charge sensing with a proximal single electron/hole transistor (SET/SHT) tuned to the flank of a Coulomb peak. Measurements are acquired in dilution refrigeration units at base temperatures below $100\!$ mK. In the case of Devices B and C, read-out was done by reflecting a microwave tone off an SHT ohmic in the range 100-250\! MHz. The signal couples to the device through an $LC$ matching network on the sample PCB, making the reflected signal sensitive to impedance changes caused by fluctuations in the device's hole occupancy \cite{vigneau2023probing}. The reflected signal is routed to a demodulation circuit at room temperature via a directional coupler mounted inside the cryogenic set-up. In the case of Device A, read-out is achieved through direct amplification of the sensor's current using a dual-stage SiGe heterojunction bipolar transistor (HBT), located on the sample PCB \cite{curry2015cryogenic}, which is further amplified at room temperature. The SET's source contact is DC biased through the junction's emitter, and \PG{coupled} to an AC signal applied on the SET's drain at around 100\! kHz. For further details on individual set-ups and fabrication, refer to \cite{george202412, saez2024microwave, john2024two} for devices A, B, and C, respectively.

\section{Further details for results}
\label{app:furtherdata}
The three test sets corresponding to devices A, B, and C contain 67, 72, and 60 CSDs, respectively. The simulated test set includes 128 CSDs.

\bibliography{bibliography}

\end{document}